\documentclass[a4paper,12pt]{article}
\usepackage{graphicx,amsmath,amssymb,rotate,textcomp,gensymb}
\usepackage{mathrsfs,float}
\usepackage{fixltx2e}
\usepackage{caption,subcaption}
\usepackage[none]{hyphenat}
\usepackage{hyperref}

\begin{document}

\title{Effect of dipolar-angle on phospholipid assembly}

\author{Tanay Paul$^1$ and Jayashree Saha$^2$\\ \small{Department of Physics, University of Calcutta,}\\ \small{92, A. P. C. Road, Kolkata - 700009, India.}\\ \small{$^1$email: tanaypaul9492@gmail.com}\\ \small{$^2$email: jsphy@caluniv.ac.in}}

\date{}

\maketitle

\begin{abstract}
We report the effect of lipid head-group dipole orientation on phase behaviour of phospholipid assembly. The work explains molecular-scale mechanism of ion-lipid, anesthetic-lipid interactions where reorientation of dipoles play important role in membrane potential modification. Molecular Dynamics simulations are performed to analyse structure-property relationship and dynamical behaviour of lipid biomembranes considering coarse-grained model interactions.
\end{abstract}

\begin{center}
 \textbf{I. Introduction}
\end{center}

\hspace{1.0cm}The complex phase behaviour of bio-membrane is indispensable for its important intra and intercellular functions which include partition, transport and communication revealing instantaneous physiological state \cite{cevc}. The structure of membrane is normally sheet-like lipid bilayer matrix with proteins embedded in it. These proteins mainly function as pumps, ion channels, receptors, transducers, enzymes. Membrane lipids, which frame the basic membrane structure, are amphiphilic having a charged or neutral hydrophilic head-group connected with two hydrophobic acyl chains. Though water plays important role in membrane functioning, the membrane structure is primarily driven by the bulky lipid because of their much slower rate of diffusive motion compared to water molecules \cite{saiz}. Lipid arrangement in membrane is fluid but highly structured, both transversely and laterally, in space and time, because of their liquid crystalline nature. Structure and dynamics of lipid assembly are becoming key objects in drug research as effective designing of new drugs and drug delivery requires insight into the physical properties of bio-membrane \cite{mouritsen}. Polar character of neutral phospholipids comes largely from the charge seperation between phosphorus($P$) and nitrogen($N$) groups in the head-group. Experimental studies indicate that the orientation of the lipid $P\rightarrow N$ vectors at room temperature are not uniformly distributed, instead, on average, their preferred orientation is along the plane of the layer \cite{saiz}. Electrostatic potential across membrane that controls membrane function mainly arises from specific preferential orientation of head-group dipoles. However, presence of ions and other charged species like drugs, anesthetics largely affects this potential by changing dipolar orientations \cite{reigada}.

\hspace{1.0cm}Shift in dipolar tilt due to rearrangement of headgroup conformation responds as charge sensor and acts as a voltmeter. It has been found that salts strongly influence the forces acting between stacks of bilayers in multilamellar vesicles. For monovalent ion types $Na^{+}$, $Cl^{-}$ etc. local ion-induced head-group tilt perturbations were found. $Na^{+}$ effectively pushes the headgroup down towards the bilayer plane as a result of the electrostatic repulsion between choline group in the lipid and cation \cite{sachs}. A nearly equal and opposite effect is seen with anion $Cl^{-}$ which pulls the head-group out. Larger anions however may swing $P\rightarrow N$ dipole towards bilayer plane as $Na^{+}$ by binding in a second deeper binding site gives two opposing effects on head-group tilt. Lipid head-group orientation is highly dependent on the specific ion present and their concentration. In pure DPPC system the $P\rightarrow N$ dipoles were preferentially oriented at an angle $\theta \approx 78\degree$ with respect to bilayer normal. In the same system with $100$ $Ca^{++}$ ions the head-group peak angle were shifted to $\theta \approx 43\degree$. The effect is less at lower concentration of $Ca^{++}$ ions \cite{vernier}.

\hspace{1.0cm}Molecular mechanism of anesthetics is a long debated issue \cite{reigada}. Two hypotheses are presently in conflict: One hypothesis \cite{franks} is based on direct action of anesthetics with specific receptor sites in integral and peripheral proteins, because experimental fact is that many anesthetics bind to specific sites; the other \cite{frankslieb} favours non-specific interaction caused by change in lipid matrix physical properties as specific binding concept is not applicable in case of a wide diversity of anesthetic compounds \cite{reigada} \cite{vernier} \cite{meyer}. The second hypothesis is supported by the Meyer-Overtones's rule \cite{meyer} \cite{overton} used for long time to correlate the potency of anesthetics and their dissolving ability in olive oil. Main problem with experimental study on this issue is that much higher concentration than clinical criterion is required to observe nerve conduction blocking \cite{lieb}. Inspite of extensive experimental and theoretical studies on the influence of local and general anesthetics on membrane structure and dynamics, there is still no clear consensus as anesthetic compounds include diverse chemical structures like small halogenated agents, alcohols with different chain lengths, bulky steroid compounds, hormones, lacking obvious structure-property relationship.

\hspace{1.0cm}Anesthetic-lipid interaction draws great interest due to several reasons. Anesthetic drug has to cross several membranes including blood-brain barrier to get access to the central nervous system, therefore, primary target is cell membrane lipid matrix. Recently, renewed attention in the role of membrane lipids are becoming prevalent by Cantor's observation \cite{cantor97} that alternation of lateral pressure profile of lipid membrane occurs with incorporation of anesthetic drugs. Pressure change modifies the opening/closing action of ion channels. With the increase in local lateral pressure channel opening requires more work, therefore the protein conformational equilibrium favours a closed state \cite{cantor99}. Thus head-group dipoles act as controlling factor for openning or closing of ion channels.

\hspace{1.0cm}Recent progress in experiments and simulations related to biomembrane study indicates combination of two conflicting hypotheses responsible for anesthetic action. It has been suggested that anesthetic action is not related to increase in lateral diffusion caused by decrease in acyl chain order as was previously assumed \cite{alakoskela}. It has been observed that electrostatic potential across membrane is primarily responsible for anesthetic action. The potential, often referred to as bilayer dipole potential, arises from the specifically oriented lipid dipoles and water dipoles at the interface. Presence of ions and charged species like anesthetics and other drugs affects this electrostatic potential significantly, by changing the dipolar orientation of lipid molecules \cite{hogberg}. It has been shown that local anesthetic articaine causes increase of dipole electrostatic potential in the membrane interior \cite{mojumdar}.

\hspace{1.0cm}Hogberg et al \cite{hogberg} showed dipolar orientation changes to $72\degree$ from $79.8\degree$ with the introduction of $12$ charged Lidocaine, a local anesthetic, and $60\degree$ with $36$ Lidocaine. Alakoskela et al \cite{alakoskela} indicated general anesthetic drugs like Pregnanolane, Isoflurane, Halothane could influence the dipolar orientation and membrane potential as well.

\hspace{1.0cm}Moreover, the study on organohalogens, chloroform and carbon tetrachloride showed \cite{reigada} that though both of them passed Meyer-Overton's criteria, the later lacks anesthetic character. The main difference between many anesthetics and their respective non-anesthetic counterparts is that the former possess dipole moment and they accumulate near the head-group regions with preferential orientation whereas non-anesthetic counterpart favour positioning at inner non-polar region comprising of lipid hydrocarbon chains. Non-immobilizer does not influence the orientation of the lipid head-group dipole moment in contrast to anesthetics \cite{koubi}.

\hspace{1.0cm}Head-group tilt is a very important influencing factor as dipole moment of the head-group is one of the main contributing factor in the membrane electrostatic potential, but the molecular scale relevance of orientation change to the structure and dynamics of membrane is not understood. Despite this immense importance, to the best of our knowledge, effect of dipolar orientation on lipid arrangement, phase behaviour and dynamics was not studied. In order to appreciate the aspect of lipid matrix properties with this effective molecular-scale structural change due to ions and drugs, we have performed coarse-grained molecular dynamics study of model lipid systems.

\begin{center}
 \textbf{II. Model and simulation details}
\end{center}
 
%\hspace{1.0cm}The anisotropic potential that is responsible for liquid crystalline mesophase has contributions both from attractive and repulsive interactions. For our simulation of dipolar molecules we use computer efficient single-site ellipsoidal potential to take into account the attractive and repulsive part of the interaction. The electrostatic part of the potential is basically a dipole-dipole interaction term. Probably the simplest and most widely used single-site ellipsoidal liquid crystal model incorporating attractive and repulsive part is the Gay-Berne potential. We have used the Gay-Berne potential with permanent dipoles to simulate bilayer. 

\hspace{1.0cm}As all-atom simulation requires a huge amount of computation time, coarse-grained generic models are used to perform semiquantative lipid simulations \cite{cooke} \cite{arnarez} \cite{marrink} \cite{farago} \cite{sodt}. Whitehead et al \cite{whitehead}, Ayton et al \cite{ayton} used G-B ellipsoid to model large length scale properties of lipid bilayers. Sun and Gezelter \cite{sun} observed ripple phases considering model lipids each consisting of a dipolar lipid head-group and ellipsoidal tail.

\begin{figure}[htp]
\centering
%\resizebox{85mm}{!}{\includegraphics {model.jpg}}
%\resizebox{85mm}{!}{\includegraphics {model2.jpg}}
\includegraphics[scale=0.35]{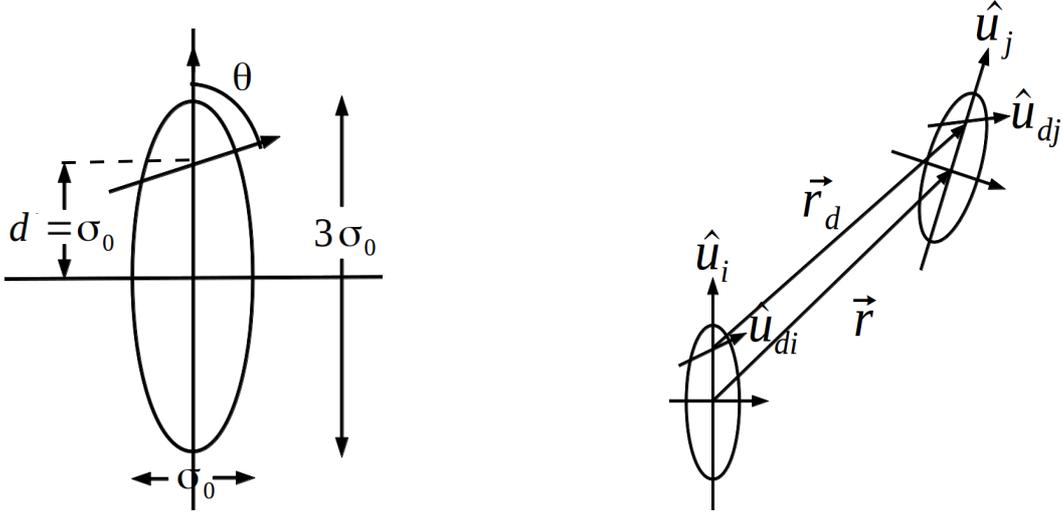}
\caption{Modeling of lipid molecule with dipolar head-group and nonpolar ellipsoidal tail}
\label{fig:modeling}
\end{figure}

\hspace{1.0cm}In our model, zwitterionic lipid molecules are considered as ellipsoidal molecules embedded with terminal point dipole. The anisotropic tail part of the lipid molecule has been modelled as prolate ellipsoid using Gaussian overlap potential. As the polar head group of the lipid molecules have dipolar character, it is modelled as simply a point dipole placed at $d=\sigma_{0}$ from the center of mass of the molecule along the molecular long axis, where $\sigma_0$ is the minor axis length of the ellipsoid (Figure: \ref{fig:modeling}). The angle $\theta$ between the symmetry axis of the molecule and the dipolar direction is an important indicator of the type of phase that will develop.

\hspace{1.0cm}The electrostatic interaction acting between the polar head-group is the interaction important for bilayer phase formation and in our model it is simply the dipole-dipole interaction. The dipole-dipole interaction potential acting between two point dipoles can be written as,
\begin{equation}
U_{dd}=\frac{1}{r^3_{d}}\left[\vec{\mu}_{d_{i}}.\vec{\mu}_{d_{j}}-\frac{3}{r^{2}_{d}}(\vec{\mu}_{d_{i}}.\vec{r}_{d})(\vec{\mu}_{d_{j}}.\vec{r}_{d})\right] \label{eq:1}
\end{equation}
where ${\bf r_{d}}={\bf \hat{r}_{d}}r_{d}$ is the vector joining the two point dipoles at distance $r_{d}$ embedded on molecules $i$ and $j$. $\vec{\mu}_{d_{i}}\equiv \mu^*{\bf \hat{u}_{d_{i}}}$, $\vec{\mu}_{d_{j}}\equiv \mu^*{\bf \hat{u}_{d_{j}}}$ are the dipole moment vectors of the point dipoles embedded on molecules $i$ and $j$ respectively. Here $\mu^*\equiv (\mu^2/\varepsilon_{s}\sigma_{0}^3)^{1/2}$ is the dimensionless dipole moment of each molecule. The long-range nature of the dipolar interaction has been taken into account considering Reaction Field method \cite{onsager}. The field on a dipole consists of two parts, a short range contribution from the molecules within a cut-off sphere of radius $r_{c}$, and molecules outside the cut-off sphere are considered to form a dielectric continuum of permittivity $\epsilon_{s}$ producing a reaction field within the cavity. The magnitude of the reaction field acting on any molecule inside the spherical cavity of radius $r_{c}$ is proportional to the total dipole moment $\sum_{j\in R }\vec{\mu_{j}}$ of the cavity due to all the molecules inside the cavity $R$ and can be expressed as
\[\vec{\varepsilon_{i}}=  \frac{2(\epsilon_{s}-1)}{2\epsilon_{s}+1}\frac{1}{r_{c}^{3}}\sum_{j\in R }\vec{\mu_{j}}  \]
The contribution to energy of $i^{th}$ molecule from Reaction Field is $-\frac{1}{2}\vec{\mu_{i}}.\vec{\varepsilon_{i}}$ and hence to energy of the system of $N$ molecules is,
\[ -\frac{1}{2}\frac{2(\epsilon_{s}-1)}{2\epsilon_{s}+1}\frac{1}{r_{c}^{3}}\sum_{i=1}^{N}\sum_{j\in R }\vec{\mu_{i}}.\vec{\mu_{j}} \]

\hspace{1.0cm}The interaction between the anisotropic tail parts of two particles $i$ and $j$ which we consider as ellipsoids of revolution is Gay-Berne potential, a most widely used single-site ellipsoidal potential incorporating attractive and repulsive parts. This G-B potential is basically an overlap potential representing van-der-Waals type interaction \cite{gb}. Taking the major axis as the molecular $z$-axis with orientation given by unit vectors ${\bf \hat{u}_{i}}$ and ${\bf \hat{u}_{j}}$ with respect to the lab frame and with centers of mass separated by $\vec{r}$, the G-B potential interacting between $i^{th}$ and $j^{th}$ particle can be written as \cite{gb},
\begin{eqnarray}
U_{GB} &=& U\left({\bf \hat{u}_{i}},{\bf \hat{u}_{j}},{\bf \hat{r}}\right) \nonumber \\
       &=& 4\varepsilon\left({\bf \hat{u}_{i}},{\bf \hat{u}_{j}},{\bf \hat{r}}\right)\left \{\left[\frac{\sigma_{0}}{r-\sigma\left({\bf \hat{u}_{i}},{\bf \hat{u}_{j}},{\bf \hat{r}}\right)+\sigma_{0}}\right]^{12}-\left[\frac{\sigma_{0}}{r-\sigma\left({\bf \hat{u}_{i}},{\bf \hat{u}_{j}},{\bf \hat{r}}\right)+\sigma_{0}}\right]^6\right \} \label{eq:2}
\end{eqnarray}
where ${\bf \hat{r}}$ is a unit vector along the intermolecular separation vector. The orientation dependent range parameter $\sigma$ is given by,
\begin{equation}
\sigma\left({\bf \hat{u}_{i}},{\bf \hat{u}_{j}},{\bf \hat{r}}\right)=\sigma_{0}\left \{1-\frac{\chi}{2}\left[\frac{({\bf \hat{u}_{i}}.{\bf \hat{r}}+{\bf \hat{u}_{j}}.{\bf \hat{r}})^2}{1+\chi({\bf \hat{u}_{i}}.{\bf \hat{u}_{j}})}+\frac{({\bf \hat{u}_{i}}.{\bf \hat{r}}-{\bf \hat{u}_{j}}.{\bf \hat{r}})^2}{1-\chi({\bf \hat{u}_{i}}.{\bf \hat{u}_{j}})}\right]\right \}^{-{\frac{1}{2}}} \label{eq:3}
\end{equation}
where $\chi$ is determined by shape anisotropy, $\kappa \equiv \left(\frac{\sigma_{e}}{\sigma_{0}}\right)$ of the particles,
\begin{equation}
\chi=\frac{\kappa^2-1}{\kappa^2+1} \label{eq:4}
\end{equation}
Here $\sigma_{e}$, $\sigma_{0}$ are size parameters reflecting the length and the breadth of the particles. The energy term in eqn.\ref{eq:2} can be written as,
\begin{equation}
\varepsilon\left({\bf \hat{u}_{i}},{\bf \hat{u}_{j}},{\bf \hat{r}}\right)=\varepsilon_0\varepsilon'^\mu\left({\bf \hat{u}_{i}},{\bf \hat{u}_{j}},{\bf \hat{r}}\right)\varepsilon^\nu\left({\bf \hat{u}_{i}},{\bf \hat{u}_{j}}\right) \label{eq:5}
\end{equation}
where, $\varepsilon_0$ is the energy scaling parameter and
\begin{equation}
\varepsilon\left({\bf \hat{u}_{i}},{\bf \hat{u}_{j}}\right)=\left[1-\chi^2\left({\bf \hat{u}_{i}}.{\bf \hat{u}_{j}}\right)^2\right]^{-{\frac{1}{2}}} \label{eq:6}
\end{equation}
and
\begin{equation}
\varepsilon'\left({\bf \hat{u}_{i}},{\bf \hat{u}_{j}},{\bf \hat{r}}\right)=1-\frac{\chi'}{2}\left[\frac{\left({\bf \hat{u}_{i}}.{\bf \hat{r}}+{\bf \hat{u}_{j}}.{\bf \hat{r}}\right)^2}{1+\chi'\left({\bf \hat{u}_{i}}.{\bf \hat{u}_{j}}\right)}+\frac{\left({\bf \hat{u}_{i}}.{\bf \hat{r}}-{\bf \hat{u}_{j}}.{\bf \hat{r}}\right)^2}{1-\chi'\left({\bf \hat{u}_{i}}.{\bf \hat{u}_{j}}\right)}\right] \label{eq:7}
\end{equation}
The parameter $\chi'$ reflects the anisotropy in the attractive forces,
\begin{equation}
\chi'=\frac{1-\kappa'^{\frac{1}{\mu}}}{1+\kappa'^{\frac{1}{\mu}}} \label{eq:8}
\end{equation}
where $\kappa'$ is the anisotropy ratio: $ \kappa' = \frac{\varepsilon_{e}}{\varepsilon_{s}} $ and $\varepsilon_{e}$, $\varepsilon_{s}$ are the well depths for the end-to-end and side-by-side configurations. The parameters $\kappa=3$, $\kappa'=1/5$, $\mu=1$, $\nu=2$ were considered in our study.

\hspace{1.0cm}The pair potential then can be written as the sum of the Gay-Berne term and the dipole-dipole term:\[U_{ij}=U_{GB}+U_{dd}\]

\hspace{1.0cm}To study effect of dipolar orientation we studied systems consisting model lipids each having specific dipolar angles. The angles between the electric dipole moment vector and the molecular orientation vector, $\theta$, are chosen with values 0\textdegree (longitudinal dipole), 30\textdegree, 45\textdegree, 50\textdegree, 55\textdegree, 60\textdegree, 75\textdegree, 90\textdegree (transverse dipole), 105\textdegree, 120\textdegree, 135\textdegree, 150\textdegree, 180\textdegree and for each angle simulation run are performed starting from isotropic phase. Density $\rho^*\left(\rho^*\equiv\frac{N\sigma_{0}^3}{V}\right)$ is set as $0.25$ for system sizes $N=500$ molecules.

\hspace{1.0cm}We use the $NVT$ Molecular Dynamics which incorporates canonical ensemble truly. A Leap-Frog algorithm for Damped Force method \cite{evans} for constatnt temperature molecular dynamics proposed by Brown and Clarke \cite{brown} has been used. For every system simulation run has been started from well equilibrated isotropic phase at $T^*=5.0$ ($T^*\equiv\frac{k_{B}T}{\varepsilon_0}$ where $k_B$ is the Boltzmann constant and $T$ is the actual temperature), then temperature has been decreased gradually to realize ordered phases and at each temperature the configuration obtained from the previous higher temperature has been used as the starting configuration. At each temperature step, system has been allowed to equilibrate keeping temperature constant for $10^5$ steps in isotropic phases, $3\times10^5$ steps in nematic phases and a long run of $5\times10^6$ steps has been performed near a transition from nematic phase to let the system acquire its stable layered phases. Simulation box volume has been set to be fixed at $V^*\equiv V/\sigma_{0}^3=1.0$.

%\newpage

\begin{center}
 \textbf{III. Simulation results}
\end{center}
 
\begin{figure}[h!]
%\resizebox{90mm}{!}{\includegraphics {qmgaconf_trans.eps}}
\begin{subfigure}{0.3\textwidth}
 \includegraphics[width=\textwidth]{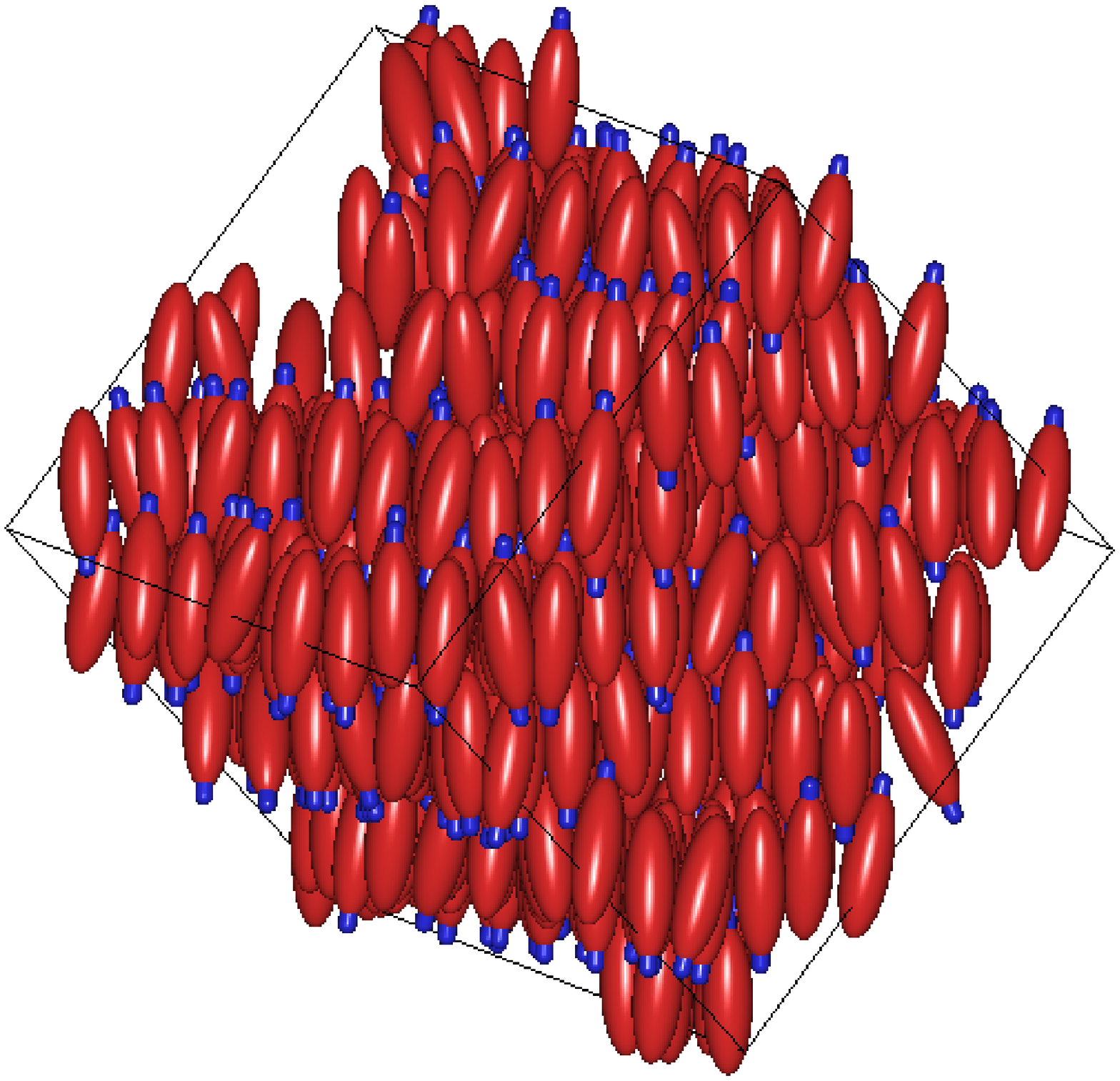}
 \caption{}
 \label{fig:0}
\end{subfigure}
\hfill
\begin{subfigure}{0.3\textwidth}
 \includegraphics[width=\textwidth]{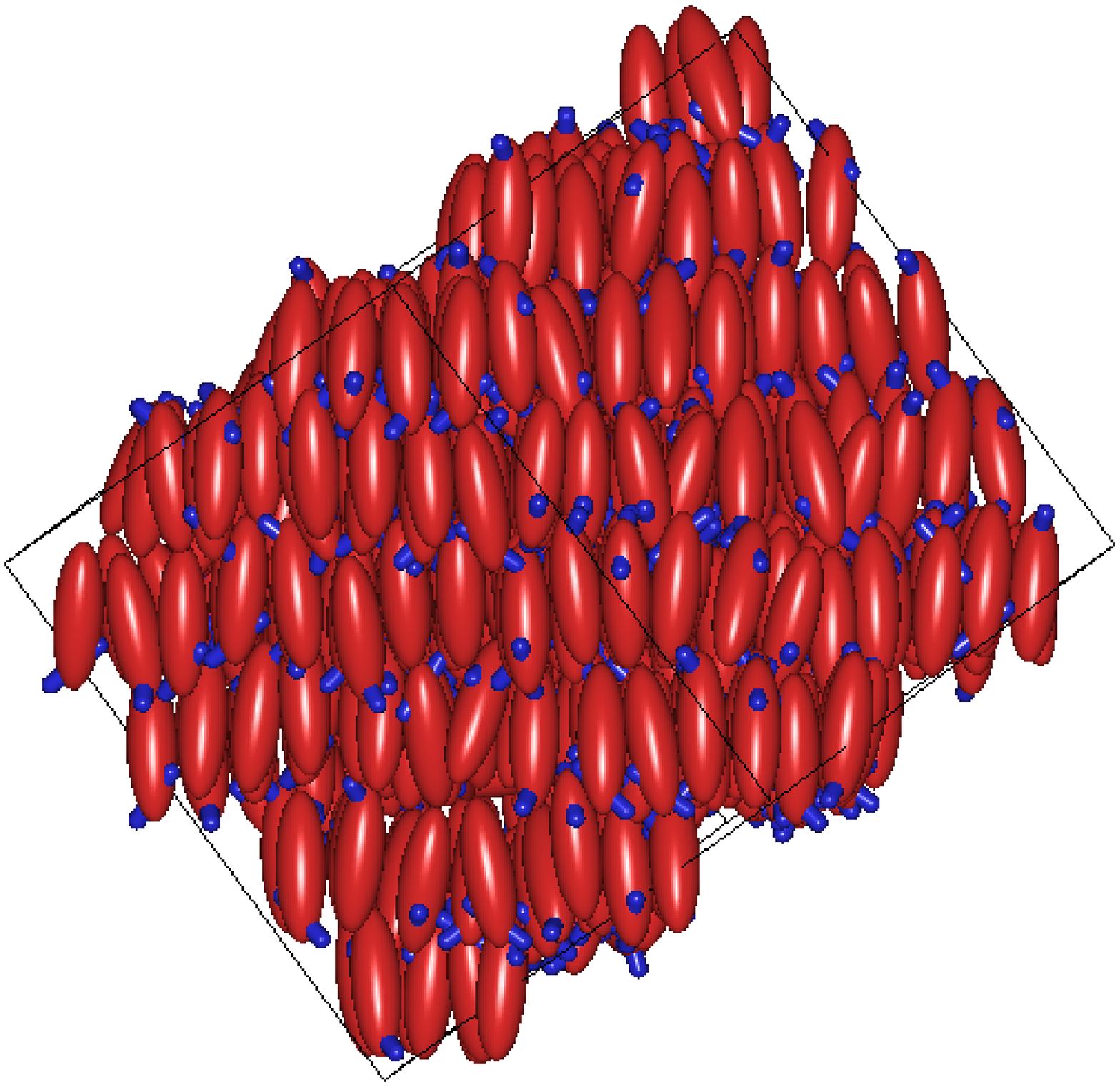}
 \caption{}
 \label{fig:45}
\end{subfigure}
\hfill
\begin{subfigure}{0.3\textwidth}
 \includegraphics[width=\textwidth]{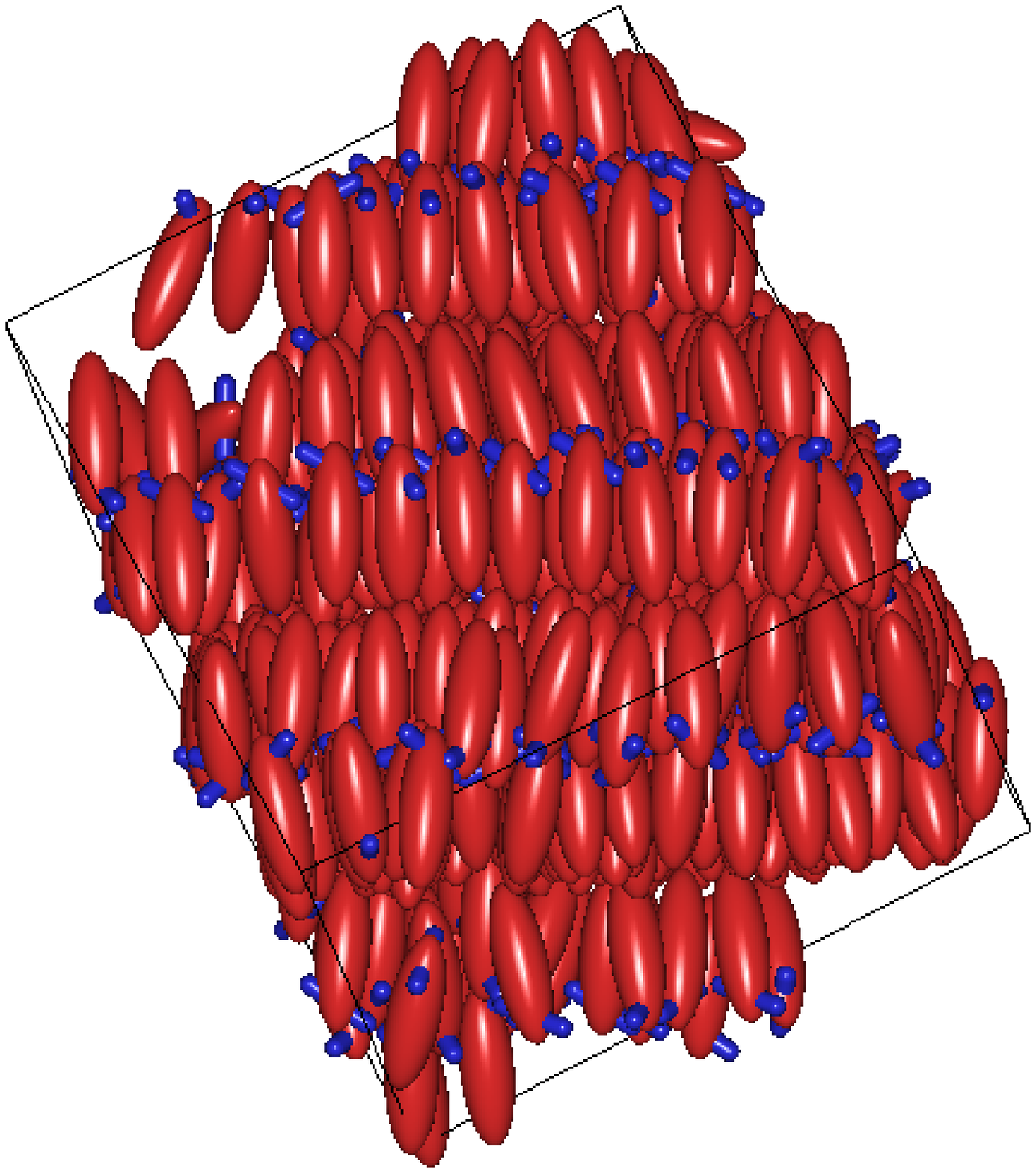}
% \vspace{0.1cm}
 \caption{}
 \label{fig:60}
\end{subfigure}
\\
%\begin{subfigure}{0.3\textwidth}
% \includegraphics[width=\textwidth]{qmgaconf_ang75.eps}
% \caption{}
% \label{fig:75}
%\end{subfigure}
%\hfill
\begin{subfigure}{0.3\textwidth}
 \includegraphics[width=\textwidth]{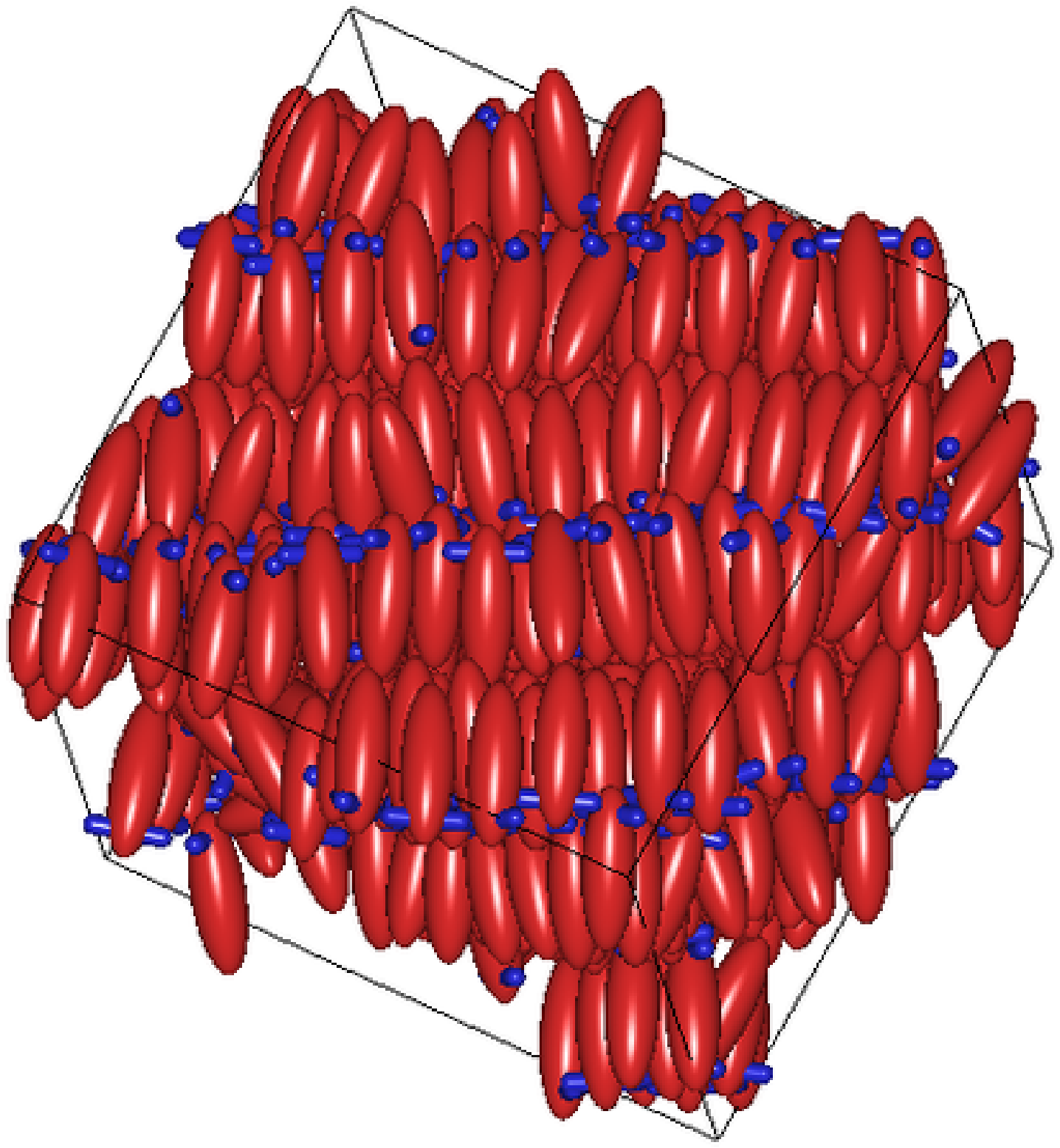}
 \caption{}
 \label{fig:90}
\end{subfigure}
\hfill
\begin{subfigure}{0.3\textwidth}
 \includegraphics[width=\textwidth]{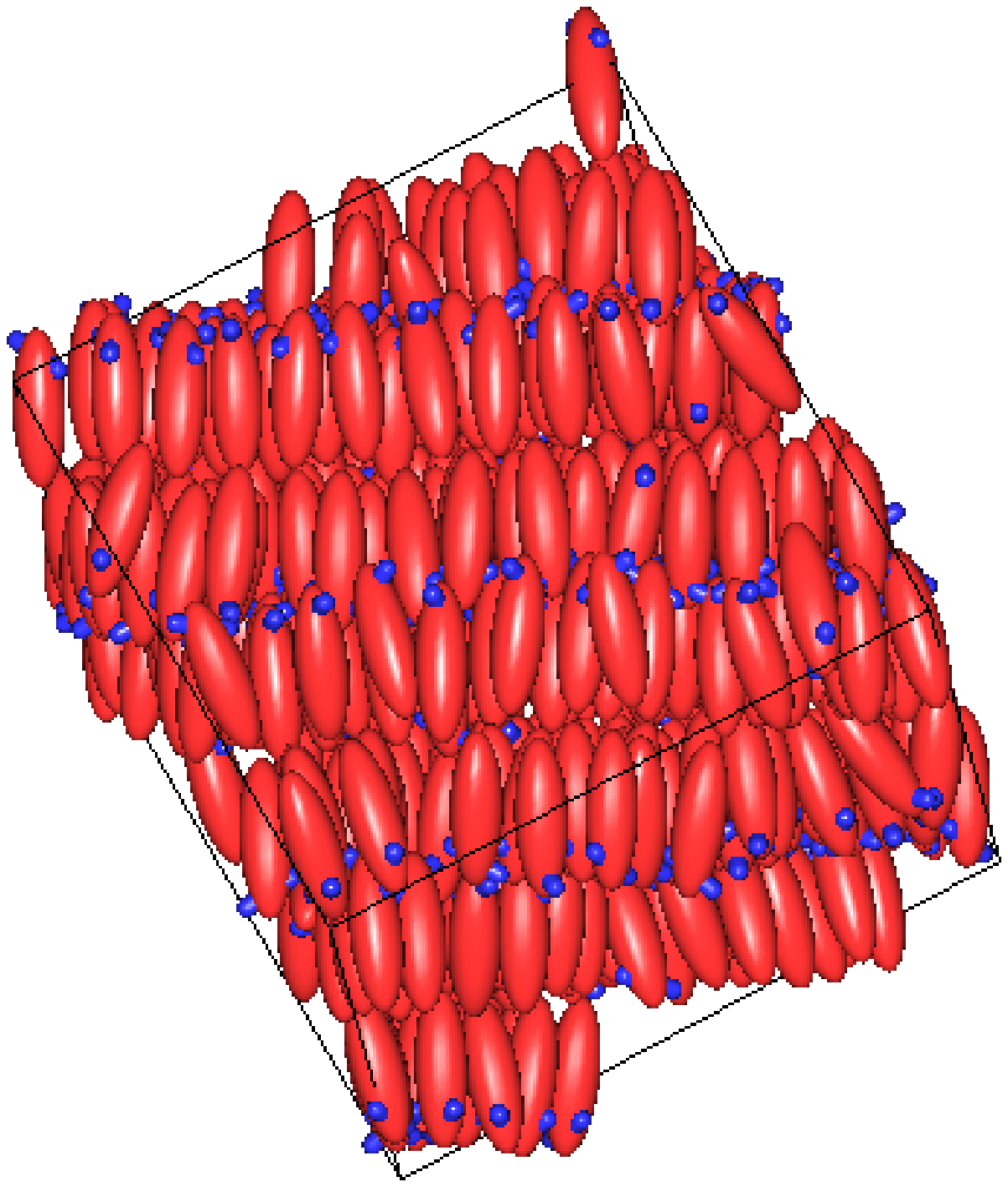}
% \vspace{0.1cm}
 \caption{}
 \label{fig:120}
\end{subfigure}
%\\
%\begin{subfigure}{0.3\textwidth}
% \includegraphics[width=\textwidth]{qmgaconf_ang120.eps}
% \caption{}
% \label{fig:120}
%\end{subfigure}
%\hfill
%\begin{subfigure}{0.3\textwidth}
% \includegraphics[width=\textwidth]{qmgaconf_ang135.eps}
% \caption{}
% \label{fig:135}
%\end{subfigure}
%\hfill
\begin{subfigure}{0.3\textwidth}
 \includegraphics[width=\textwidth]{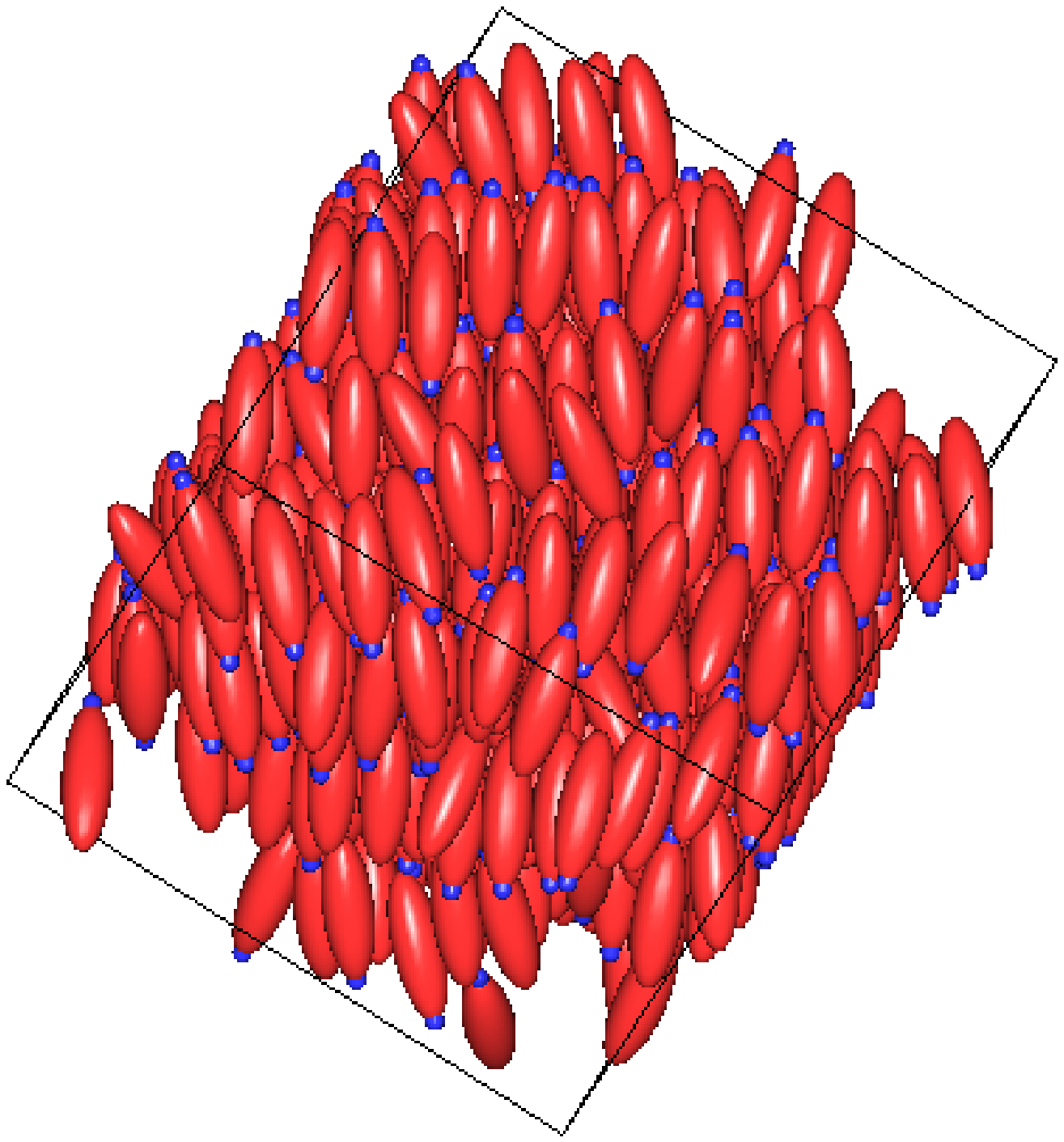}
% \vspace{0.1cm}
 \caption{}
 \label{fig:180}
\end{subfigure}
\caption{Snapshots of the configurations generated by MD simulation for $N=500$: (a)$\theta=0\degree$ at $T^*=0.7$,(b)$\theta=45\degree$ at $T^*=0.9$,(c)$\theta=60\degree$ at $T^*=1.0$,(d)$\theta=90\degree$ at $T^*=1.1$,(f)$\theta=120\degree$ at $T^*=1.0$,(g)$\theta=180\degree$ at $T^*=0.8$. Molecules are shown in red and dipoles in blue.}
\label{fig:layeredconfig}
\end{figure}

\hspace{1.0cm}The length to breadth ratio of molecules is chosen as $\kappa \equiv (\frac{\sigma_{e}}{\sigma_{0}})=\frac{3}{1}$, where $\sigma_{e}$ and $\sigma_{0}$ are size parameters reflecting the major axis length and the minor axis length of the ellipsoid. Table \ref{table:1} shows stable phases formed for different dipolar orintations at different temperatures for system of $N=500$ molecules. The reduced temperatures shown in the Table \ref{table:1} are the transition temperatures at which the systems crystalizes. For all the systems with differnt dipolar angles the system was allowed to evolve from isotropic phase. Decreasing the temperature and allowing the systems to equilibrate at constant temperature for generation of layered stable phases. Snapshots of the configurations obtained at fluid layered phases for systems with $N=500$ molecules with different dipolar orientation are exhibited in figure \ref{fig:layeredconfig} . Molecules with dipoles at dipolar angles $60\degree, 75\degree, 90\degree, 120\degree$ form bilayered phases ( fig.\ref{fig:60} \ref{fig:90} \ref{fig:120} ) with density $\rho^*=0.25$ at different temperatures for systems with different dipolar angles. The corresponding transition temperature is maximum for transverse dipole and is relatively smaller for other angles. In case of smaller angles there exists flipped lamellar phases with no preferred head-tail direction and this is also true for angles greater than $120\degree$. We have performed our study for different system size $N=256$ and similar results are obtained for all system sizes. In this simulation, to reduce computation time, we have not taken into account the water interaction directly, therefore inter-bilayer gaps are not prominent. However, the dimensionless dipole moment $\mu^*$ of each molecule has been taken as $1.1$ which is relatively smaller than the actual value of the lipid head-group dipole moment. Smaller magnitude of dipole moment but having same value for all the systems with different dipolar angles is considered to bring out the dipolar angle effect on phase behaviour qualitatively.
%\begin{figure}[t!]
%\centering
%\includegraphics[scale=0.35]{qmgaconf_transiso.eps}
%\caption{A typical isotropic phase for the system with $\theta=90\degree$ and $N=500$}
%\label{fig:isotropic}
%\end{figure}

\begin{table}[h!]
 \centering
 \begin{tabular}{|c|c|c|c|c|c|c|c|c|c|c|c|}
 \hline
 $\theta$ & 0\textdegree & 30\textdegree & 45\textdegree & 50\textdegree & 55\textdegree & 60\textdegree & 75\textdegree & 90\textdegree & 120\textdegree & 135\textdegree & 180\textdegree \\ \hline
 Phases & L\textsubscript{F} & L\textsubscript{F} & L\textsubscript{F} & L\textsubscript{F} & L\textsubscript{F} & B & B & B & B & L\textsubscript{F} & L\textsubscript{F} \\ \hline
% $\theta$\\ \hline
% Phases \\ \hline
% 256 & 0.25 & 0.7 & 0\textdegree & lamellar\\ \cline{3-5}
%  & & 0.7 & 30\textdegree & lamellar\\ \cline{3-5}
%  & & 0.8 & 45\textdegree & lamellar\\ \cline{3-5}
%  & & 0.9 & 50\textdegree & lamellar\\ \cline{3-5}
%  & & 0.9 & 55\textdegree & antiphase\\ \cline{3-5}
%  & & 0.9 & 60\textdegree & bilayer\\ \cline{3-5}
%  & & 0.9 & 75\textdegree & bilayer\\ \cline{3-5}
%  & & 1.0 & 90\textdegree & bilayer\\ \cline{3-5}
%  & & 0.9 & 105\textdegree & antiphase\\ \cline{3-5}
%  & & 0.9 & 120\textdegree & bilayer\\ \cline{3-5}
%  & & 0.9 & 135\textdegree & lamellar\\ \cline{3-5}
%  & & 0.7 & 180\textdegree & lamellar\\ \hline
  \end{tabular}
\caption{Stable phases formed with variation of dipolar angle with molecular long axis for $500$ molecules and $\rho^*=0.25$. Here `L\textsubscript{F}' stands for `Flipped Lamellar' and `B' stands for `Bilayer'}
\label{table:1}
\end{table}

\begin{figure}[h!]
\centering
\begin{subfigure}{0.3\textwidth}
 \includegraphics[width=\textwidth]{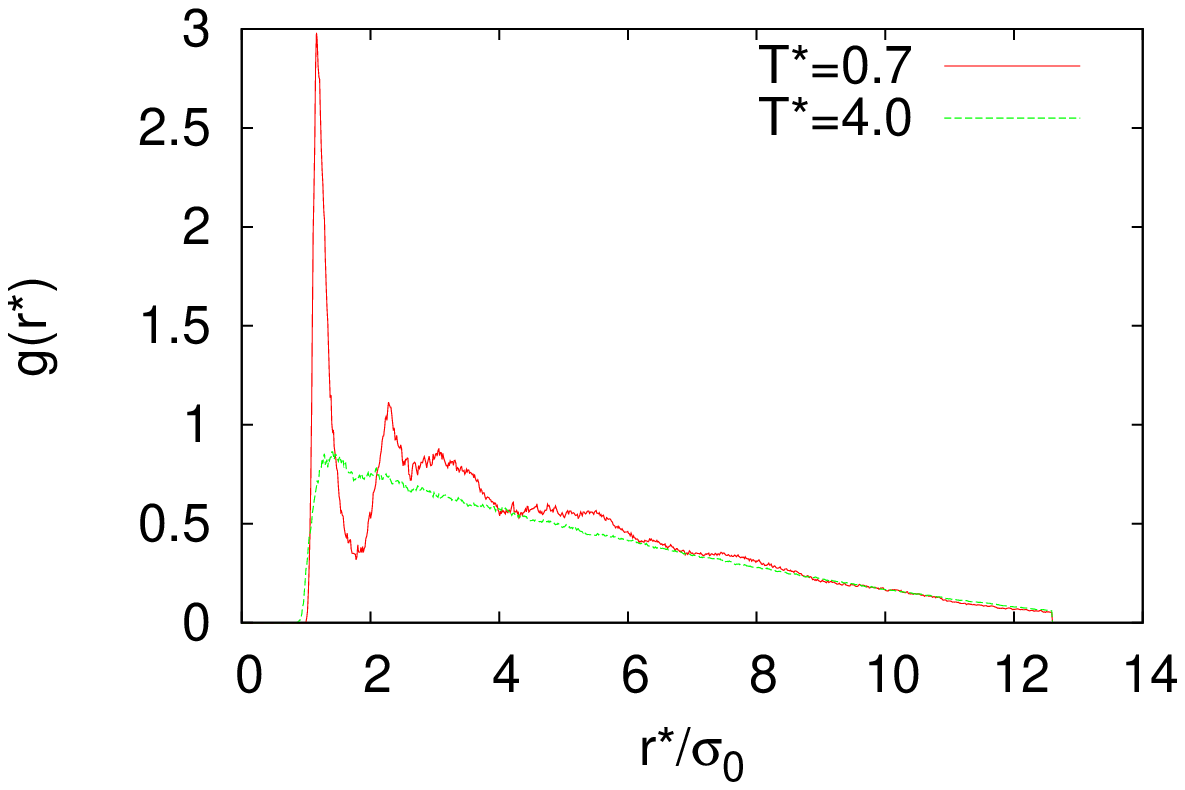}
 \caption{}
 \label{fig:gr0}
\end{subfigure}
\hfill
\begin{subfigure}{0.3\textwidth}
 \includegraphics[width=\textwidth]{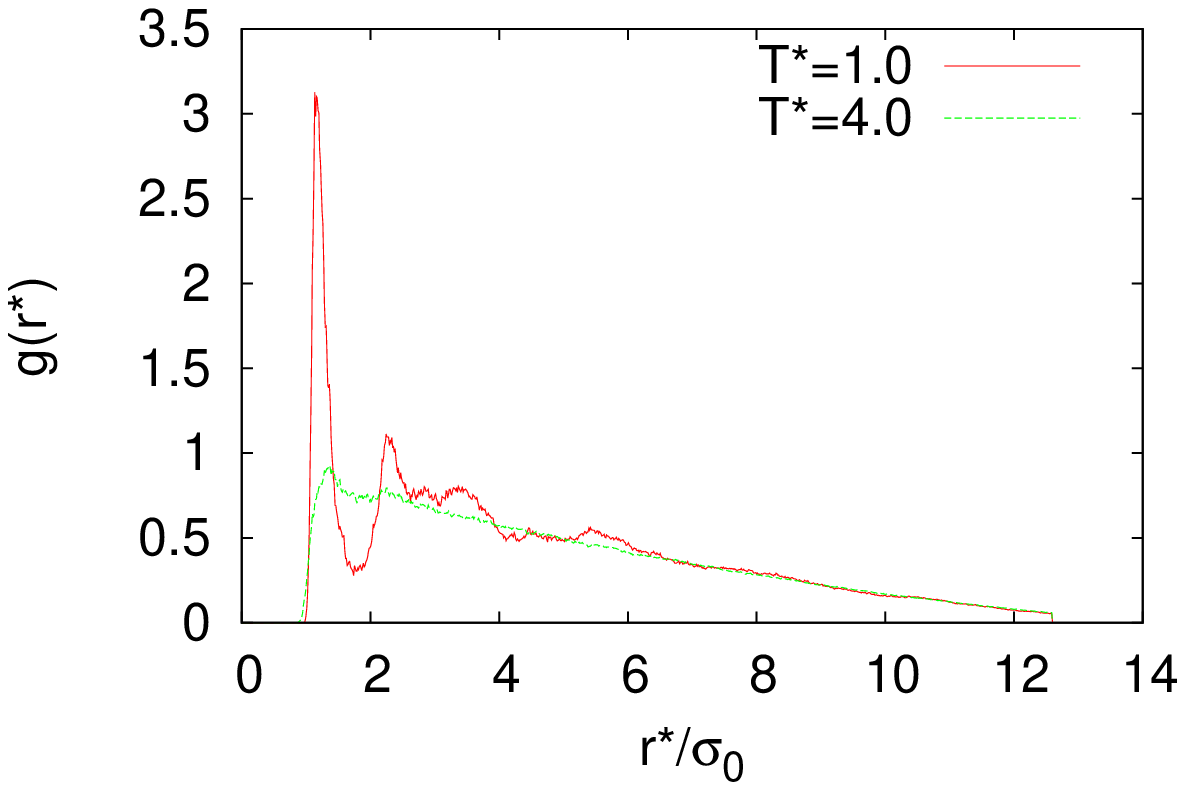}
 \caption{}
 \label{fig:gr60}
\end{subfigure}
\hfill
\begin{subfigure}{0.3\textwidth}
 \includegraphics[width=\textwidth]{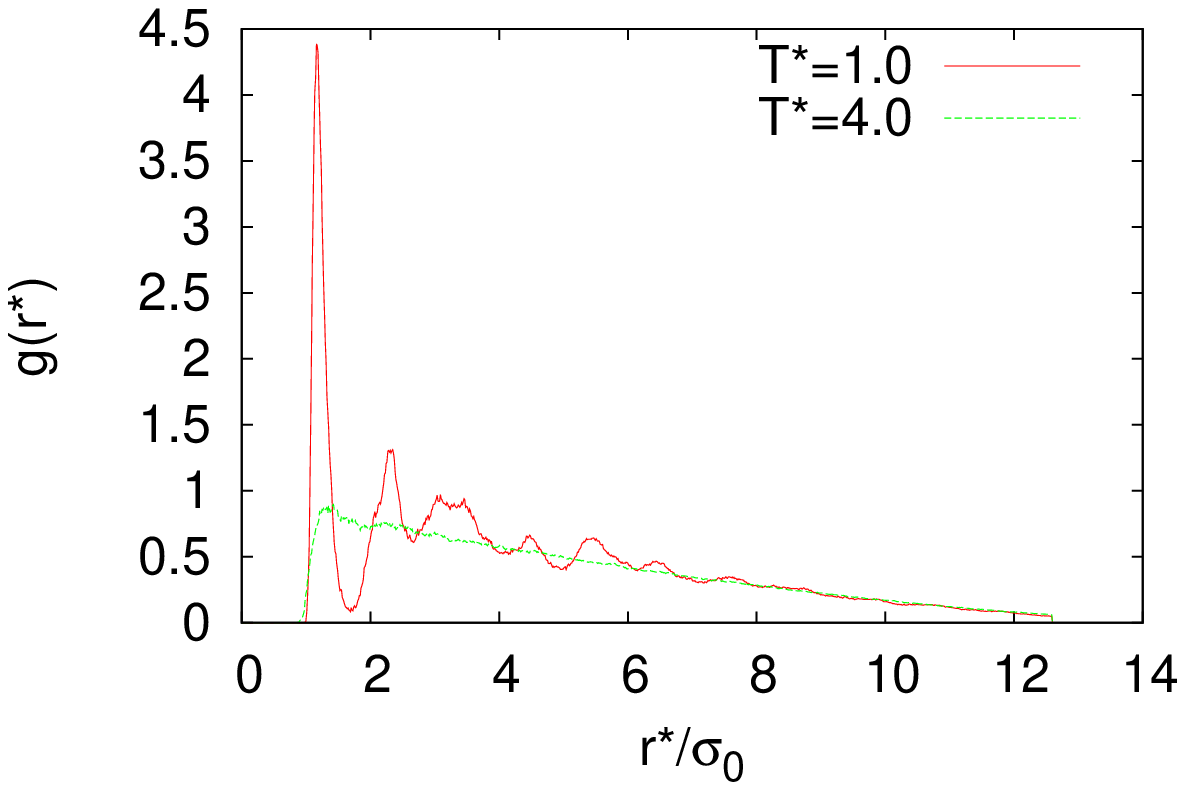}
 \caption{}
 \label{fig:gr90}
\end{subfigure}
\caption{Comparison of $g(r^*)$ for systems with different angles at stable layered phases and in a typical isotropic phase. (a) for $\theta=0\degree$, (b) for $\theta=60\degree$, (c) for $\theta=90\degree$}
\label{fig:gr}
\end{figure}

\begin{figure}[ht!]
\centering
 \includegraphics[width=0.85\textwidth]{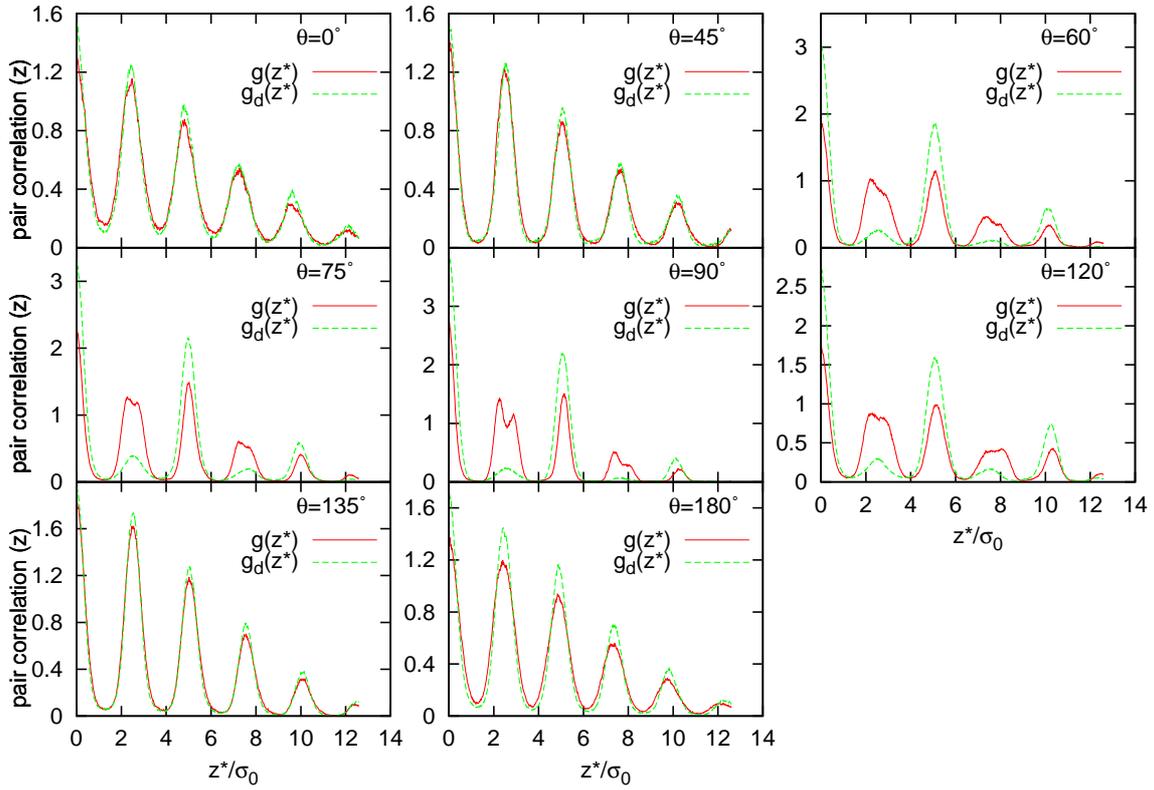}
 \caption{Plots of $g(z^*)$ and $g_d(z^*)$ for systems with different dipolar angles}
 \label{fig:gzvsgdz}
\end{figure}

\hspace{1.0cm}For structural analysis some distribution functions have been calculated. We plotted pair distribution function or radial distribution function $g(r^*)$ (fig.\ref{fig:gr}), pair correlation function of center of masses of molecules along director axis $g(z^*)$ and dipolar positions $g_{d}(z^*)$ (fig.\ref{fig:gzvsgdz}), where $r^*$ is the seperation between two molecular center of masses and $z^*$ is the length of the projection of the seperation vector between two molecular center of masses in case of $g(z^*)$ and two dipolar positions in case of $g_{d}(z^*)$ in reduced unit. The plots of $g(z^*)$ and $g_{d}(z^*)$ shows the existence of peaks for dipolar correlation function at the alternate peaks of center of mass correlation function along director axis which clearly indicates the presence of bilayer at $\theta = 90\degree, 75\degree, 60\degree$ and $120\degree$. But for the systems with other angles peaks for both the functions co-exist indicating non-existence of bilayered phases. The plot of $g(r^*)$ shows existence of some translational order locally but not globally referring to fluidity.

\hspace{1.0cm}The results of our study clearly shows the existence of bilayered lipid phases for dipolar angles greater than and equal to $60\degree$ and less than and equal to $120\degree$. For other angles lamellar phases show non bilayer structures. At some angles less than $60\degree$ and greater than $55\degree$ it appears that antiphase structure can exist and at smaller angles though flipped layered structures exist the bilayer phase formation does not occur. Same is true for angles greater than $120\degree$.

\begin{figure}[hb!]
\centering
\begin{subfigure}{0.4\textwidth}
 \vspace{0.7cm}
 \includegraphics[width=\textwidth]{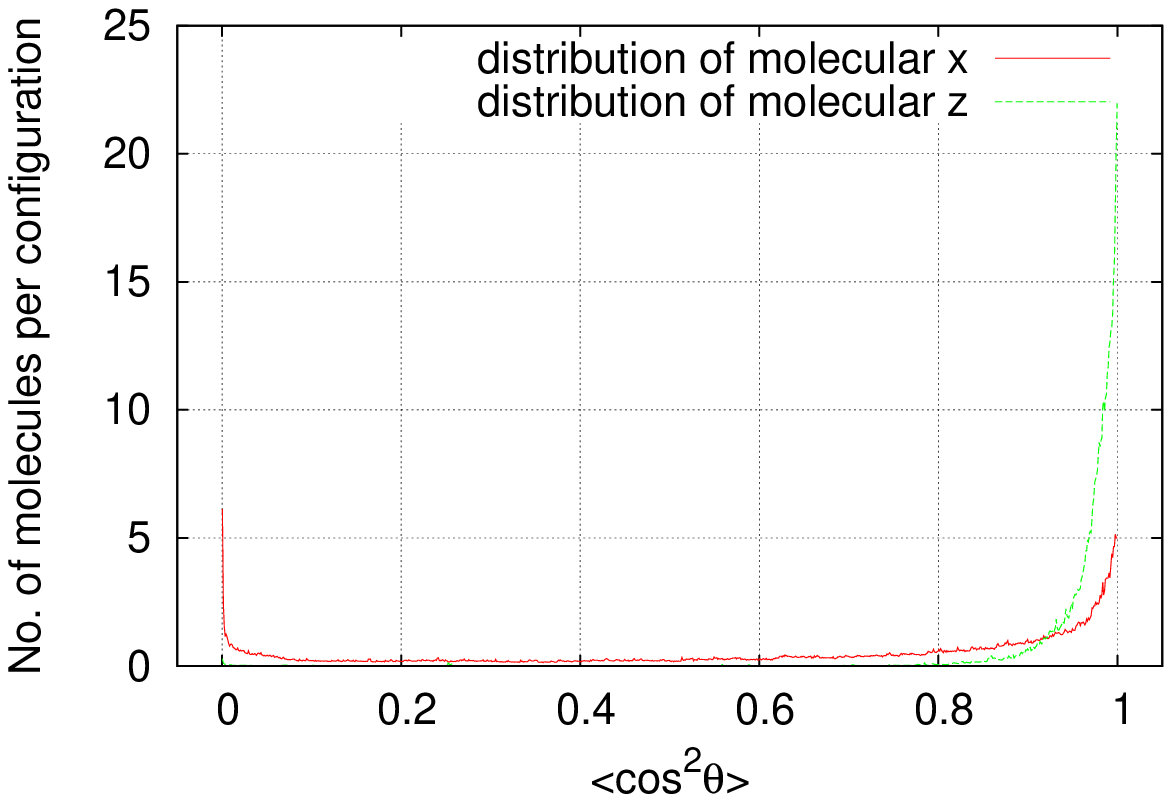}
 \vspace{0.7cm}
 \caption{}
 \label{fig:xyrot_trans}
\end{subfigure}
\hspace{0.5cm}
\begin{subfigure}{0.4\textwidth}
 \includegraphics[width=\textwidth]{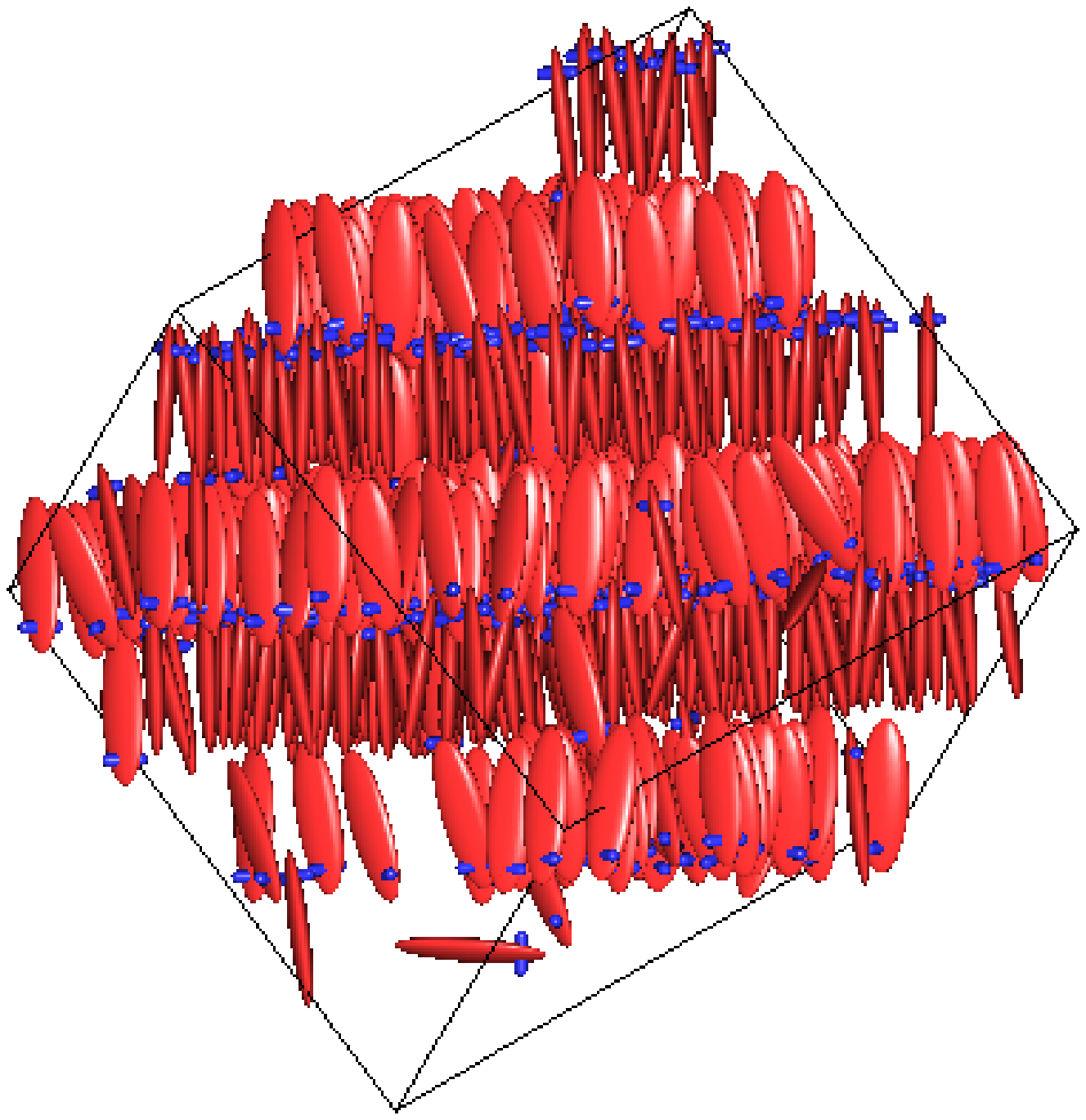}
 \caption{}
 \label{fig:xyrotcnf_trans}
\end{subfigure}
\caption{(a)Distribution of molecular x-axis and z-axis, (b)Configuration for $\theta=90\degree$ with planes containing molecular x-axis and dipoles. Molecules are shown in red and dipoles in blue.}
\label{fig:cossqrth_trans}
\end{figure}

\hspace{1.0cm}Another interesting observation of the study is that in the bilayer phase molecules of neighbouring layers are rotated about the z-axis with respect to each other whereas it vanishes in non-bilayer phase. Snapshot ( fig. \ref{fig:xyrotcnf_trans} ) of planes containing molecular x-axis as well as dipoles shows that respective molecular planes of each layer are rotated making angles $\approx 90\degree$ with respect to the molecules belonging to the adjacent layers. The distribution of molecular z-axes with respect to the molecular director axis has been plotted in green and distribution of molecular x-axes with respect to most probable molecular x-direction has been plotted in red ( fig. \ref{fig:xyrot_trans} ). The interaction between molecules are effectively biaxial in nature because of dipolar parts. Biaxiality is maximum for $\theta=90\degree$ i.e. for transverse dipoles and decreases with deviation in $\theta$ from $90\degree$ and vanishes at $\theta=0\degree$ or $180\degree$.

%\newpage

\begin{center}
 \textbf{IV. Conclusion}
\end{center}
 
\hspace{1.0cm}We performed molecular dynamics simulations for systems comprising of single component lipid molecules having different dipolar angles. For the realization of bulk phase behaviour, we considered coarse-grained lipid model to simulate large-scale systems. The simulation results reported in this work are important in respect to gaining insight about physiological ion balancing and rational drug designing because underlying molecular scale mechanism responsible for structural and dynamical rearrangement of lipid lamellar phases upon addition of salt, anesthetics, other charges are not directly available from experiments or continuum descriptions. Solvent interactions are not taken into account directly in this simulation work as our aim is to report results providing only qualitative picture of head-group dipolar orientation effect on the behaviour of liquid crystalline layered phases, as a result of electric field modification. With the introduction of drugs and ions, changes in head-group dipolar orientation occur. Deviation of dipolar angle from $\theta\approx 90\degree$ actually weakens the bilayer formation ability of the lipid assembly. This instability increases with larger reorientation and random flipping of molecules destroy bilayer structure outside $120\degree<\theta<60\degree$ range. The lipid layers were unable to retain bilayer arrangement for head-group dipoles with angles $\theta<60\degree$ and $\theta>120\degree$ but having same dipole moment. The study also provides molecular-scale insight to the phenomena that head-group dipolar preferential orientation in nature is along the layer planes because this favours bilayer structure of bio-membrane most. Moreover, this characteristic change in layer structure can be used to design efficient liposomes in such a way that it can fuse with biomembranes at suitable conditions \cite{safinya}. This change in liquid crystalline phase behaviour with head-group dipolar orientation may be explored to achieve considerable impact on liposomal drug and gene delivery which requires membrane fusion at certain stage of action.

\begin{center}
 \textbf{V. Acknowledgement}
\end{center}
 
\hspace{1.0cm}T.P. gratefully acknowledges the support of Council of Scientific {\&} Industrial Research (CSIR), India, for providing Junior Research Fellowship. This work is supported by the UGC-UPE scheme of the University of Calcutta.

\end{document}